\documentclass{mem}
\usepackage{natbib}\usepackage{txfonts}\usepackage{balance}
\usepackage{graphicx}
\usepackage[a4paper,breaklinks,dvipdfm]{hyperref}
\idline{1}{1}

\begin{document}
\def\teff{$T\rm_{eff }$}
\def\kms{$\mathrm {km s}^{-1}$}

\title{
Gravitational Waves optical follow-up at VST
}

   \subtitle{}

\author{
A. \,Grado\inst{1} 
\and E. \,Brocato\inst{2} 
\and M. \,Branchesi\inst{3}
\and E. \,Cappellaro\inst{4} 
\and S. \,Covino\inst{5}
\and M. \, Della Valle\inst{1}
\and F. \,Getman\inst{1}
\and G. \,Greco\inst{6,7}
\and L. \,Limatola\inst{1}
\and G. \,Stratta\inst{6,7}
\and S. \,Yang\inst{4} \\
on behalf of the larger GRAWITA collaboration \inst{}
} 

\institute{
INAF, Osservatorio Astronomico di Capodimonte, salita Moiariello 16, I-80131, Napoli, Italy 
\and INAF, Osservatorio Astronomico di Roma, Via di Frascati, 33, I-00078 Monteporzio Catone, Italy
\and Gran Sasso Science Institute,  Viale F. Crispi, 7 67100 L'Aquila, Italy
\and INAF, Osservatorio Astronomico di Padova, Vicolo dell'Osservatorio 5, I-35122 Padova, Italy
\and INAF, Osservatorio Astronomico di Brera, Via E. Bianchi 46, I-23807 Merate (LC), Italy
\and Universit\`a degli Studi di Urbino `Carlo Bo', Dipartimento di Scienze Pure e Applicate, P.za Repubblica 13, I-61029, Urbino, Italy
\and INFN, Sezione di Firenze, I-50019 Sesto Fiorentino, Firenze, Italy\\
\email{aniello.grado@.inaf.it} 
}

\authorrunning{Grado }

\titlerunning{GW follow-up at VST }

\abstract{
We report on the deep optical follow-up surveys of the first two 
gravitational-wave events, GW150914 and GW151226, accomplished by the 
GRAvitational Wave Inaf TeAm Collaboration (GRAWITA) using the VLT 
Survey Telescope (VST). We responded promptly to the gravitational-wave 
alerts sent by the LIGO and Virgo Collaborations, covering a region of  
$90$ deg$^2$ and $72$ deg$^2$ for GW150914 and GW151226, respectively, 
and kept observing the two areas over nearly two months. Both surveys reached 
an  average limiting magnitude of about 21 in the $r-$band. The paper 
outlines the VST observational strategy and two independent procedures 
developed to search for transient counterpart candidates in multi-epoch 
VST images. Numerous transients have been discovered, mostly variable stars 
and eclipsing binaries, but no candidates 
are identified as related to the gravitational-wave events. 
The work done let on to gain experience and tune the tools for next LVC runs 
and in general to exploit the synergies between wide field optical surveys and 
future multi-messenger programs including projects like Theseus.  

\keywords{Multi-messenger, Gravitational Waves, Wide field astronomy}
}
\maketitle{}

\section{Introduction}

Gravitational waves (GWs) are perturbations of space-time metric produced by a time dependent  mass quadrupole moment. GWs are emitted  by different kinds of astrophysical sources. Among those, coalescence of binary systems of compact objects such as two neutron stars (BNS), NS and a stellar-mass black hole (NSBH) or two black holes (BBH), collapse of massive stars with large degree of asymmetry and fast rotating asymmetric isolated NSs.

{In September 2015, the longstanding search for gravitational radiation was finally achieved with the detection by the LIGO and Virgo Collaboration (LVC) of unambiguous emission of GW radiation from an astrophysical source. After detailed analysis, it was recognized that the emission was originated by the coalescence of two BHs at a cosmological redshift of $z\simeq{}0.09$ \citep{Abbott2016a}. After two months, at the end of December 2015,  the GWs emitted by a second BBH system, again at $z\simeq{}0.09$, was detected \citep{Abbott2016d}. The sky localization of the two GW sources, obtained through triangulation with the two aLIGO interferometers located in Hanford (Washington) (H1) and Livingston (Louisiana) (L1) \citep{Aasi2015}, spans from a few hundreds to thousand of square degrees \citep[][]{Singer2014,Essick2015}. The large sky error box is the major challenge for the search and identification of possibly associated electromagnetic (EM) counterparts.

In this paper we describe the observational campaign performed by the GRAvitational Wave INAF TeAm (GRAWITA) \footnote{https://www.grawita.inaf.it/} to follow up the first two GW triggers during the first LVC  scientific run (O1)  by using the ESO-VLT Survey Telescope (VST), its results and the prospects for the upcoming years. 
In section \ref{obs} some details on the VST telescope  and the observational strategy  are presented, including  the specific observational response to the  LVC triggers  GW150914 and GW151226. A brief summary of the adopted pre-reduction is described in section~\ref{dataproc}. In the same section, we present  our  approach to the  transient search and  introduce the two independent pipelines we developed to this aim.  In the following section~\ref{result-foll}, the results of the search are described. For each of the two GW alerts the list of transient candidates is discussed. A brief discussion close the paper (section~\ref{concl}).

\section{VST observational strategy}
\label{obs}

\begin{table*}
\parbox{.45\linewidth}{
\centering
\caption{\label{data} Log of the VST observations performed for the GW150914 event. The covered area and the night average seeing full width half maximum are reported in the last two columns.}
\begin{tabular}{lcccc}
\\
\hline\hline
 &    ~~~~~~~  GW150914 &
\\
\hline\hline
Epoch & Date & Area & FWHM\\
     &   (UT)   &    deg$^2$    &  arcsec \\
\hline
1       & 2015-09-17 & 54 & 0.9\\ 
2       & 2015-09-18 & 90 & 0.9\\ 
3       & 2015-09-21 & 90 & 0.9\\ 
4       & 2015-09-25 & 90 & 1.1\\ 
5       & 2015-10-01 & 72 & 1.0\\ 
5       & 2015-10-03 & 18 & 1.0\\ 
6       & 2015-10-14 & 45 & 1.5\\ 
6       & 2015-11-16 &  9 & 1.2\\ 
6       & 2015-11-17 & 18 & 1.1\\ 
6       & 2015-11-18 & 18 & 1.5\\ 
\hline
\end{tabular}
}
\hfill
\parbox{.45\linewidth}{
\centering
\caption{\label{data2} Log of the VST observations performed for the GW151226 event. The covered area and the night average seeing full width half maximum are reported in the last two columns.}
\begin{tabular}{lcccc}
\\
\hline\hline
 &    ~~~~~~~  GW151226 &
\\
\hline\hline
Epoch & Date & Area & FWHM\\
     &   (UT)   &    deg$^2$    &  arcsec \\
\hline
1       & 2015-12-27 & 72 & 1.0\\ 
2       & 2015-12-29 & 72 & 1.6\\ 
3       & 2015-12-30 & 9 & 1.3\\ 
3       & 2016-01-01 & 45 & 0.9\\ 
3       & 2016-01-02 & 9 & 0.9\\ 
4       & 2016-01-05 & 18 & 1.2\\ 
4       & 2016-01-06 & 18 & 1.1\\ 
4       & 2016-01-07 & 27 & 0.8\\ 
5       & 2016-01-13 & 45 & 1.5\\ 
5       & 2016-01-14 & 27 & 1.1\\ 
6       & from 2016-01-28 &  & \\
        & to 2016-02-10 & 63 & 1.1\\
\hline
\end{tabular}
}
\end{table*}

The first GW candidate alert was sent by the Ligo Virgo Community (LVC) on 16 September 2015. After the real-time processing of data from H1 and L1, an event occurred on 14 September 2015 at 09:50:45 UTC was identified \citep[][]{GCN18330}. 

Further analysis showed that the GW event was produced by the coalescence of two black holes with rest frame masses of $29^{+4}_{-4}$M$_{\odot}$ and $36^{+5}_{-4} $M$_{\odot}$ at a luminosity distance of  $410^{+160}_{-180}$ Mpc \citep{Abbott2016b}.
On 26 December 2015, a further GW candidate (GW151226) was observed by LVC  \citep[][]{GCN18728}.
Again, the GW event resulted from the coalescence of two black holes of rest frame masses of $14.2^{+8.3}_{-3.7}$ M$_{\odot}$ and $7.5\pm2.3$ M$_{\odot}$
at a distance of $440^{+180}_{-190}$ Mpc \citep{Abbott2016d}. The multi-messenger follow-up started on 27 December 2015, more than 1 day after the GW trigger \citep[][]{GCN18728}, again with an excellent response from astronomers' community.

For the search of possible associated optical transients, our team exploited the ESO VST, a 2.6m, 1 deg$^2$ field of view (FoV) imaging telescope located at the Cerro Paranal Observatory in Chile \citep{Capaccioli2011,Kuijken2011} and dedicated to large sky surveys in the austral hemisphere.  The telescope optical design allows to achieve a uniform PSF with variation $<4\%$ over the whole field of view. 
The VST is equipped with the OmegaCAM camera, which covers the field of view of 1 square degree with a scale of 0.21 arcsec/pixel, through a mosaic of 32 CCDs.

The required time allocation was obtained in the framework of the Guarantee Time Observations (GTO) assigned by ESO to the telescope and camera teams in reward of their effort for the construction of the instrument. 
The planned strategy of the follow up transient survey foresees to monitor a sky area of up to 100 deg$^2$ at 5/6 different epochs beginning soon after the GW trigger and lasting 8-10 weeks.

With the announcement of each trigger, different low-latency probability sky maps\footnote{FITS format files containing HEALPix (Hierarchical Equal Area isoLatitude Pixelization) sky projection, where to each pixel is assigned the probability to find the GW source in that position of the sky.} were distributed to the teams of observers \citep[][]{GCN18728, GCN18330}. For GW150914  two initial sky maps of 310 deg$^2$ and 750 deg$^2$ (90 \% confidence) were produced by un-modelled  searches for GW bursts, one by the coherent Wave Burst (cWB) pipeline \citep{Klimenko2016} and the other by the Bayesian inference algorithm LALInferenceBurst (LIB) \citep{Essick2015}, respectively. For GW151226, the initial localization was generated by the Bayesian  localization algorithm BAYESTAR \citep{Singer2016} encompassing a 90\% confidence region of 1400 deg$^2$.
To prepare the Observing Blocks  (OBs) we used a dedicated script named \emph{GWsky} \footnote{https://github.com/ggreco77/GWsky}.

The typical VST OB contains groups of nine pointings (tiles) covering an area of 
$3 \times 3\; \rm{deg}^2$. For  each pointing, we obtained two exposures of 40\,s each dithered by $\sim 0.7-1.4$ arcmin to fill the OmegaCAM CCD mosaic gaps. The surveys of both events were performed in the $r$ band filter. Summary of the VST follow-ups of GW 150914 and 151226 are reported in Tab.\,\ref{data} and\,\ref{data2}, respectively.
The VST responded promptly to the GW150914 alert by executing six different OBs on 17th of September, $23$ hours after the alert  \citep{GCN18336}. Monitoring of the  region was repeated \citep{GCN18397} over two months  for a total of six observation epochs.
The VST observations captured a containment probability of 29\%. This value dropped to 10\% considering the LALinference sky map, which was shared with observers on 2016 January 13 \citep[][]{GCN18858} and showed a 90\% confidence region slightly different with respect to the first distributed skymap.. Prompt response, survey area and depth make a unique combination of features of our VST survey  matched only by the DECam survey \citep{DECam}.
Also the response to GW151226 was rapid, 7.6 hours after the alert and 1.9 days after the merger event \citep{GCN18734}. Eight OBs covered $72$ deg$^2$ corresponding to the most probable region of the GW signal visible by VST and with an airmass smaller than $2.5$. Like for GW150914, the GW151226 survey consists of 6 epochs, spanning over one and a half month.
For GW151226 the VST observations captured a containment probability of 9\% of the initial BAYESTAR sky map and 7\% of the LALinference sky map, which was shared on January 18 \citep{GCN18889} and covers a 90\% credible region of $1240$ deg$^2$.

\section{Data Processing}
\label{dataproc}

\subsection{Pre-reduction}

Immediately after acquisition, the images are mirrored to ESO data archive, and then  transfered by an automatic procedure from ESO Headquarters to the VST Data Center in Naples. The first part of the image processing was performed using {\tt VST-tube}, which is the pipeline developed for  the VST-OmegaCAM mosaics \citep{Grado12}.  It includes pre-reduction, astrometric and photometric calibration and mosaic production. For further details on the data reduction see \cite{capaccioli2015}.

\subsection{Transient search}

In order to search for variable and transient sources, the images were analysed by using two independent procedures. One based on the comparison of the photometric measurements of all the sources in the VST fields obtained at different epochs. The second is based on the analysis of the difference of images
following the approach of the supernova (SN) search program recently completed with the VST \citep{Botticella2016}. 

The two approaches are intended to be complementary, with the first typically more rapid and less sensitive to image defects and the latter more effective for sources projected over extended objects or in case of strong crowding. 
For both procedures, the main goal of our analysis is to identify sources showing a ``significant" brightness variation (around 0.5 mag ), either raising  or declining flux, during the period of monitoring, that can be associated to extra-galactic events.

The photometric pipeline (ph-pipe) is intended to provide a list of transients in low-latency to organize immediate follow-up activities. The computation time can be particularly rapid, e.g. just a few minutes for each epoch VST surveyed area. The weakness of this approach is that sources closer than about a Point Spread Function (PSF) size or embedded in extended objects can be difficult to detect and therefore can possibly remain unidentified. The procedure has been coded in the {\tt python}
~(version 3.5.1) language making use of libraries part of the {\tt anaconda}\footnote{https://docs.continuum.io/anaconda/index}~(version 2.4.1) distribution.  The procedure includes a number of basic tools to manage the datasets, i.e. source extraction, classification, information retrieval, mathematical operations, visualization, etc. Data are stored and managed as {\tt astropy}\footnote{http://www.astropy.org}~(version 1.2.1) tables. For further details see \citep{Brocato2018}

The second (ph-diff) approach is based  on 
a widely used, most effective method for transient detection, i.e. the difference of images taken at different epochs. To implement this approach for the survey described in this paper we developed a dedicated pipeline exploiting our experience with the medium-redshift SN search done with the VST \citep[SUDARE project,][]{Cappellaro15}. The pipeline is a collection of {\tt python} scripts that include specialized tools for data analysis, e.g. {\tt SExtractor}\footnote{http://www.astromatic.net/software/sextractor} \citep{BertinSEX} for source extraction and {\tt topcat}\footnote{http://www.star.bris.ac.uk/~mbt/topcat/}/{\tt stilts}\footnote{http://www.star.bris.ac.uk/~mbt/stilts/} for catalog handling. For optical images taken from the ground, a main problem is that the PSF is different at different epochs, due to the variable seeing. The PSF match is secured by the {\tt hotpants}\footnote{http://www.astro.washington.edu/users/becker /v2.0/hotpants.html} code \citep{Becker2015}, an implementation of the \citet{Alard99} algorithm for image analysis. For further details see \citep{Brocato2018}. The image difference pipeline was definitely more time consuming than the photometric pipeline, and optimization of the code is on going.

A comparison between the transients identified by the two pipelines shows that, as expected, the  image-difference pipeline is more effective, in particular for objects very close to extended sources. However, the photometric pipeline is less affected by image defects as halos of very bright or saturated stars, offering a profitable synergy. Typically, a percentage ranging from 80 to 90\% of the transients identified with the photometric pipeline are also recorded by the image-difference pipeline.

\section{Results}
\label{result-foll}

We now know that both the gravitational wave events considered here, GW150914 and GW151226, were generated by coalescence of black-holes. In the current scenario strong electromagnetic radiation is not expected to occur, and in fact none of the transients identified by the worldwide astronomical effort could be linked to the observed GW events. However, the analysis of the data obtained in response to the GW triggers is important both for evaluating the search performances and for tuning future counterpart searches. 

\subsection{GW150914}\label{G15}

The list of variable/transient objects selected by the {\tt diff-pipe} consists of ~8000 sources, of those 6722 are known variables, while the number of sources provided by the {\tt ph-pipe} is 939 (evident spurious and known variables already removed). More than 90\% of them are also detected by the {\tt diff-pipe}. The smaller number of sources detected by the {\tt ph-pipe} is due to $i)$ the removal of all the ``bright" and/or previously known variable sources after the match with the GAIA catalog and $ii)$ the much higher adopted detection threshold. Most of the sources identified by the {\tt ph-pipe} and not included in the catalog produced by the {\tt diff-pipe} turned out to be real and were typically located in regions that needed to be masked for a reliable image subtraction.
Many of the {\tt diff-pipe} candidates are known variables. As a further test, we applied the same selection criteria of the {\tt ph-pipe} to the list of the 33787 variable/transient sources identified by {\tt diff-pipe}. The selection produces a list of about 3000 objects. This last sample still includes known variable sources (more than 400) or objects whose light-curves can be classified with known templates, or possible defects in the subtraction procedure. As expected, the {\tt diff-pipe} is more effective in finding variable/transient objects than the {\tt ph-pipe}, although the final cleaned lists also contain objects that are found by one pipeline only.
A cross-check of our {\tt diff-pipe} candidate catalog against the SIMBAD database gave a match for 6722 objects of which 6309 identified with different type of variable sources, mainly RRLyrae (48\%), eclipsing  binaries (23\%) and a good number of Long Period Variables, semi-regular and Mira (23\%). The sky distribution of the matched sources reflects the LMC coverage by both our and the OGLE surveys. 
  
Searching the list of known SNe\footnote{We used the update version of the Asiago SN catalog \citep[http://sngroup.oapd.inaf.it/asnc.html,][]{Barbon99}}, we found that in the time window of interest for our search, three SNe  and one SN candidate were reported that are expected to be visible in our search images,  All these sources were detected in our images. In addition, we also found a few objects that most likely are previously undiscovered SNe.  Assuming all these objects are SNe and including the three other SNe first discovered in other surveys \footnote{we did not consider the likely AGN OGLE-2014-SN-094}, we count 10 SNe. This can be compared with the expected number of SNe based on the known SN rates in the local Universe, the survey area, the light curve of SNe, the time distribution of the observations, the detection efficiencies at the different epochs \citep[c.f. Sect. 5.1 of][]{Smartt16}.  For this computation we used a tool specifically developed for the planning of SN searches \citep{Cappellaro15}. We estimate an expected number of 15-25 SNe that suggest that our detection efficiency is roughly 50\%.

\subsection{GW151226}

The follow-up campaign for GW151226 was also characterized by a prompt response to the trigger and deep observations over a large sky area (see Section~\ref{obs}). Different from the follow-up campaign carried out for GW150914, the covered fields are at moderate Galactic latitude and close to the Ecliptic. In fact, the total number of analyzed sources was about an order of magnitude below the former case.

The {\tt diff-pipe} procedure produced a list of 6310 candidates of which 3127 with high score. Performing a crosscheck of our candidate catalog with SIMBAD database gave 54 matches with known variable sources.
The candidate list shows a large number of transients that appear only at one epoch due to the large contamination from minor planets,  which was expected for the  projection of the GW151226 sky area onto the Ecliptic.
A query with Skybot\footnote{http://vo.imcce.fr/webservices/} showed a match of 3670 candidates with known minor planets within a radius of 10\arcsec.  
The {\tt ph-pipe} yielded 305 highly variable/transient
sources (after removing the known sources reported in the GAIA catalogue and the
known minor planets).
90\% of them are also part of the list provided by the {\tt diff-pipe}. Even for GW151226 most of the sources identified by the {\tt ph-pipe} and not included in the catalog produced by the {\tt diff-pipe} turned out to be real.

We searched in our candidate list the sources detected by the Pan-STARRS (PS) survey from Table\,1 of  \cite{Smartt16b}. Of the 56 PS objects 17 are in our survey area. Out of these, 10 ($\sim 60$\%) were identified also
by our pipelines as transient candidates. The main reason for the missing detections is the lack of proper reference images. It is worth noting that in the ESO/VST archive we could not find exposures for the surveys area of the two triggers obtained before the GW events. Therefore, we have an unavoidable bias against the detection of transients with slow luminosity evolution in the relatively short time window of our survey.

\section{Summary.}
\label{concl}

GRAWITA contributed to the search of the optical counterparts of the first direct detections of GWs, GW150914 and GW151226, exploiting the capabilities of the VLT survey telescope. None of the transients identified by our team can be related to the gravitational events. Nevertheless, this work made possible to verify the capabilities, reliability and the effectiveness of our project. We started the VST observations within 23 hours after the alert for GW150914 \citep{GCN18330}, and 9 hours after the alert for GW151226 \citep{GCN18728}. Concerning the observational strategy, for GW150914, VST covered  $\simeq 90$  square degrees of the GW probability sky map in the $r$ band for 6 epochs distributed over a period of 50 days. The contained probability resulted to be one of the largest obtained by optical ground based telescopes reacting to the GW150914 alert \citep{Abbott2016c}. For GW151226, the GW sky maps favoured the observation sites located in the northern hemisphere, however we were able to monitor 2 probability regions (North and South) for a total area of $\simeq 72$ square degrees for a period of 40 days.
For both the alerts, a limiting magnitude  of the order of $r$ $\simeq$ $21$  mag was reached in most of the epochs.
For the search of optical transients two independent pipelines have been developed. One based on source extraction and magnitude comparison between different epochs and the second on transient identification obtained through image subtraction techniques.
The two pipelines allow the identification of a number of astrophysical transients, none of them can be related with plausibly reasons to the gravitational event GW150914 and GW151226.
Further steps toward a rapid detection and characterization of the optical transients are critical points that need to be addressed.
 
\begin{acknowledgements}
This paper is based on observations made with the ESO/VST. We acknowledge the usage of the VST Italian GTO time. We also acknowledge INAF financial support of the project "Gravitational Wave Astronomy with the first detections of adLIGO and adVIRGO experiments".
{\it Facility:} {VST ESO programs 095.D-0195, 095.D-0079 and 096.D-0110, 096.D-0141}. 
\end{acknowledgements}

\bibliographystyle{aa}

\end{document}